\documentclass[11pt]{article}

\usepackage{geometry}  
\usepackage[T1]{fontenc}  
\usepackage[utf8]{inputenc}  
\usepackage{lineno}  
\usepackage{fancyhdr}  
\usepackage{enumitem}  
\usepackage{hyperref}  
\usepackage{titling}  
\usepackage{natbib}  
\usepackage{mathtools}  
\usepackage{titlesec}  
\usepackage{lastpage}  

\usepackage[english]{babel}  
\usepackage{amsmath}  
\usepackage{amsfonts}  
\usepackage{amssymb}  
\usepackage{wasysym}  
\usepackage{bbm}  
\usepackage{array}  
\usepackage{xr}  
\usepackage{verbatim} 
\usepackage{float} 
\usepackage{caption}
\usepackage{subcaption}

\usepackage{csvsimple}
\usepackage{algorithm}
\usepackage{algpseudocode}

\newif\ifincludeplots
\includeplotstrue 

\newboolean{isarxiv}
\setboolean{isarxiv}{true}
\newboolean{doSuppressSubmittedTo}
\setboolean{doSuppressSubmittedTo}{false}

\usepackage{tikz}
\usetikzlibrary{shapes,arrows}

\newcommand{\unit}[1]{\ensuremath{\, \mathrm{#1}}}

\hypersetup{%
	pdfborderstyle={/S/U/W 1}
}

\setcitestyle{round,numbers}
\titleformat{name=\section}{\normalfont\Large\bfseries}{}{0pt}{}
\titleformat{name=\subsection}{\normalfont\large\bfseries}{}{0pt}{}
\titleformat{name=\subsubsection}{\normalfont\normalsize\bfseries}{}{0pt}{}

\newcommand{\email}[1]{\href{mailto:{#1}}{{#1}}}

\newcommand{\keywords}[1]{\textbf{Keywords}: {#1}}

\newcommand{\optincludegraphics}[2][{}]{\includegraphics[{#1}]{#2}}

\newcommand{\optinput}[1]{}

\newtagform{brackets}{[}{]}
\usetagform{brackets}

\newcommand{\thejournal}[1]{Magnetic Resonance in Medicine}

\title{Servo navigation and phase equalization enhanced by run-time stabilization (PEERS) for 3D EPI time series}

\ifthenelse{\boolean{doSuppressSubmittedTo}}{\rhead{\small}}{\rhead{\small \textsc{Submitted to \thejournal}}}
\newcommand{\pmcOn}{\emph{Servo on}}
\newcommand{\pmcOff}{\emph{Servo off}}

\newcommand{\tnav}{T_{nav}}
\newcommand{\subPeers}{\mathrm{epi}}
\newcommand{\subRef}{\mathrm{ref}}
\begin{document}

\begin{titlepage}
	{\noindent\LARGE\bf \thetitle}
	
	\bigskip
	
	\begin{flushleft}\large
		Malte Riedel\textsuperscript{1,{*}},
		Thomas Ulrich\textsuperscript{1},
		Samuel Bianchi\textsuperscript{1},
		Klaas P. Pruessmann\textsuperscript{1}
	\end{flushleft}
	
	\bigskip
	
	\noindent
	\begin{enumerate}[label=\textbf{\arabic*}]
		\item Institute for Biomedical Engineering, ETH Zurich and University of Zurich, Switzerland
	\end{enumerate}
	
	\bigskip
	
	
	\textbf{*} Corresponding author:
	
	\indent\indent
	\begin{tabular}{>{\bfseries}rl}
		Name		& Dr. Malte Riedel \\
		Department	& Institute for Biomedical Engineering \\
		Institute	& ETH Zurich and University of Zurich \\
		Address 	& Gloriastrasse 35 \\
		& 8092 Zurich \\
		& Switzerland \\
		E-mail		& \email{malteriedel@mailbox.org} \\
	\end{tabular}
	
	\vfill
	
	
    \ifthenelse{\boolean{isarxiv}}{}{
	\wcTotal
    \thispagestyle{empty}
	
	\noindent
	Abstract: 232 / 250 allowed\\
	
	\noindent
	Introduction: 587\\
	Methods: 1578\\
	Results: 1119\\
	Discussion: 1354\\
	Body total: 4638 / 5000 allowed\\
	}
\end{titlepage}

\pagebreak

\begin{abstract}{
	\noindent\textbf{Purpose}: To enhance time-resolved segmented imaging by synergy of run-time stabilization and retrospective, data-driven phase correction.
	
    \noindent\textbf{Methods}: A segmented 3D EPI sequence for fMRI time series is equipped with servo navigation based on short orbital navigators and a linear perturbation model, enabling run-time correction for rigid-body motion as well as bulk phase and frequency fluctuation. Complementary retrospective phase correction is based on the repetitive structure of the time series and serves to address residual phase and frequency offsets. The combined approach is termed phase equalization enhanced by run-time stabilization (PEERS).
    
    \noindent\textbf{Results}: The proposed strategy is evaluated in a phantom and in-vivo. Servo navigation is found to diminish motion confound in raw data and maintain k-space consistency over time series. In turn, retrospective phase equalization is found to eliminate shot-wise phase and frequency offsets relative to the navigator, which are attributed to eddy-currents and vibrations from phase encoding. Retrospective phase equalization reduces the precision requirements for run-time frequency control, supporting the use of short navigators. Relative to conventional volume realignment, PEERS achieved tSNR improvements up to $30\%$ for small motion and in the order of $10\%$ when volunteers tried to hold still. Retrospective phase equalization is found to clearly outperform phase correction based solely on navigator-based frequency estimates.
    
    \noindent\textbf{Conclusion}: Servo navigation achieves high-precision run-time motion correction for 3D EPI fMRI. Coarse frequency tracking based on short navigators is supplemented by precise retrospective frequency and phase correction. Fully automatic and self-calibrated, PEERS offers effective plug-and-play motion and phase correction for 3D fMRI.}
\end{abstract}

\bigskip

\pagebreak

\hypertarget{introduction}{%
	\section{Introduction}\label{introduction}}
Functional magnetic resonance imaging (fMRI) requires fast volume acquisitions to resolve BOLD signal variations \cite{ogawa_intrinsic_1992,ogawa_functional_1993,belliveau_functional_1991} in space and time. Due to its fast sampling and inherent robustness, two-dimensional echo-planar imaging (2D EPI) \cite{mansfield_multi-planar_1977} has become a standard technique for fMRI acquisitions. 3D EPI \cite{song_echovolume_1994,mansfield_echo-volumar_1995} offers intrinsically higher sensitivity and more benign spin history-related motion artifacts than 2D EPI, but makes volume acquisition more vulnerable to physiological noise \cite{poser_three_2010,zaitsev_prospective_2017}. Head motion and magnetic field fluctuations over the course of 3D scans contaminate image time series and statistical data analysis, posing further challenges for fMRI with 3D EPI \cite{van_der_zwaag_temporal_2012,van_dijk_influence_2012,jorge_signal_2013,lutti_highresolution_2013,tijssen_optimizing_2014,todd_prospective_2015,liu_noise_2016,reynaud_influence_2017}.

In principle, head motion can be addressed by robust sampling, motion prevention, and motion correction \cite{zaitsev_motion_2015,maclaren_prospective_2013,godenschweger_motion_2016}. Rapidly sampled volume acquisitions in fMRI are in fact fairly robust to motion, yet inter-volume motion and artifacts due to intra-volume motion can still persist \cite{zaitsev_prospective_2017}. Motion prevention techniques like dedicated patient preparation \cite{de_bie_preparing_2010}, bite bars \cite{menon_design_1997}, cushioning or sedation \cite{bernal_fmri_2012} can be used to reduce motion artifacts with varying costs and effectiveness. In addition, a wide range of prospective and retrospective techniques have been proposed to effectively correct head motion in various scan settings based on external sensor hardware \cite{zaitsev_magnetic_2006,ooi_prospective_2009,haeberlin_real-time_2015,todd_prospective_2015,schildknecht_wireless_2019,van_niekerk_toward_2019}, MR-based navigators \cite{ehman_adaptive_1989,fu_orbital_1995,welch_spherical_2002,van_der_kouwe_real-time_2006,white_promo_2010,kober_head_2011,skare_properties_2015,wallace_head_2019,liu_reducing_2020,ulrich_servo_2024}, or data-driven self-navigation \cite{atkinson_autofocus_1997,bammer_augmented_2007,loktyushin_blind_2013,graedel_motion_2017,polak_scout_2022,bayih_self-navigated_2022}.

Navigator-based motion corrections offer high-precision motion tracking without the need for external hardware, but at the cost of sequence interference \cite{maclaren_prospective_2013}. Recently, methods using short navigators of a few milliseconds have been proposed \cite{wallace_head_2019,ulrich_servo_2024,riedel_run-time_2025}, which can be easily inserted into standard fMRI sequences. The servo navigation method \cite{ulrich_servo_2024,riedel_run-time_2025} uses short navigator readouts ($\approx3\unit{ms}$) and a self-calibrating linear model for run-time motion and frequency correction.
Once implemented, such navigated sequences offer autonomous, easy-to-use operation, making motion correction more accessible for a wide range of studies.

In addition to head motion, magnetic field fluctuations are another confounding factor in image time series \cite{koch_optimization_2009,juchem_b0_2017,stockmann_vivo_2018}. There are many causes for field variations, including breathing \cite{raj_model_2000,van_de_moortele_respiration-induced_2002}, pose-dependent field changes \cite{maclaren_prospective_2013,marques_application_2005,liu_effect_2018}, scanner drift or sampling-related eddy currents \cite{boesch_temporal_1991,vannesjo_gradient_2013}, which have different properties regarding spatial complexity and time evolution. Again, sensing mechanisms include external hardware \cite{van_gelderen_real-time_2007,wilm_higher_2011,kasper_monitoring_2015,duerst_utility_2016,kasper_rapid_2018,engel_singleshot_2018,vionnet_simultaneous_2021}, MR navigators \cite{liu_reducing_2020,ulrich_servo_2024,riedel_run-time_2025,wallace_rapid_2020,splitthoff_sense_2009,wallace_realtime_2022}, and data-driven techniques \cite{splitthoff_sense_2009}, which can be used in reconstruction or at run time. 

Scans with repeated volume acquisitions as in fMRI additionally offer the opportunity to leverage redundancies along the time series not only for volume alignment \cite{liu_noise_2016,bayih_self-navigated_2022} but also for phase correction \cite{hu_retrospective_1995,wowk_k-space_1997,durand_artifact_2001,pfeuffer_correction_2002,moeller_self-navigation_2020,parker_k-space-based_2022}. 
The \emph{dynamic off-resonance in k-space} (DORK) method \cite{pfeuffer_correction_2002} provides shot-wise phase and frequency corrections (temporal first-order phase model) from a two-point phase fit between image raw data in the k-space center and a navigator echo. In contrast, self-navigation techniques monitor shot-to-shot changes in phase accrual from image raw data alone. Shot-wise phase correction of zeroth-order in time was originally demonstrated for 2D FLASH imaging \cite{wowk_k-space_1997} and has recently been extended to first-order phase correction in 3D EPI for MR thermometry in phantoms and tissue samples \cite{parker_k-space-based_2022}.
All of these techniques rely on stable k-space consistency over time and are thus sensitive to head rotations.

The goal of the present work is to improve the robustness of 3D fMRI by synergy of run-time stabilization, using servo navigation, and retrospective, data-driven phase correction as proposed by Parker et al. \cite{parker_k-space-based_2022}. Servo navigation with motion and frequency update \cite{ulrich_servo_2024} is included in  a segmented 3D EPI sequence. The navigator's frequency precision proves to be critical for robust navigator-based EPI phase correction, and dedicated correction strategies are investigated. Self-navigated retrospective correction \cite{parker_k-space-based_2022} fine-tunes per-shot phase and frequency estimates, exploiting redundancy due to repetition along the time series. The combined approach, phase equalization enhanced by run-time stabilization (PEERS), is explored in phantoms and in-vivo, with and without motion. Special attention is given to the synergy between the two strategies in comparison with their individual performance.
 

\hypertarget{theory-and-methods}{%
	\section{Methods}\label{theory-and-methods}}

\begin{figure}[H]
    \centering
	\ifincludeplots\optincludegraphics[width=\textwidth]{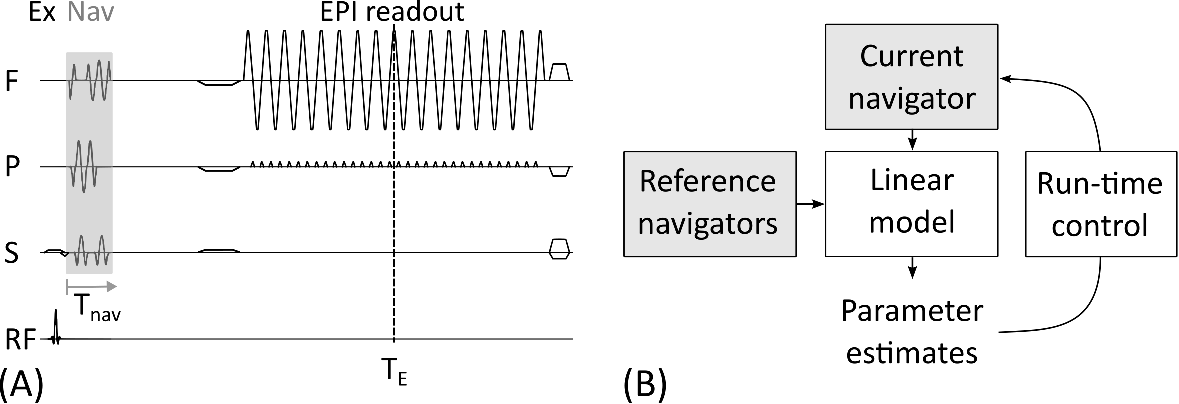}\fi
	\caption{Servo navigation for segmented 3D EPI. (A) Sequence diagram. (B) Servo navigation overview. A 3D orbital navigator (Nav) with adaptable duration $T_{nav}$ is inserted between excitation (Ex) and EPI readout. After steady-state build-up, reference navigators are acquired on the fly to calibrate the linear model for motion and field parameter estimation. The parameters are used for run-time scan geometry control closing the feedback loop for updated linear parameter estimation.}
	\label{fig:fig1_sequence_control}
\end{figure}

\hypertarget{methods-servo-navigation}{%
	\subsection{Servo navigation for segmented 3D EPI}\label{methods-servo-navigation}}
The concept of servo navigation has previously been studied for 3D gradient echo imaging \cite{ulrich_servo_2024,riedel_run-time_2025} and is, in this work, integrated into a segmented 3D EPI sequence. The sequence diagram is shown in Fig. \ref{fig:fig1_sequence_control}A and consists of a slab-selective excitation (Ex) and an EPI readout. In each EPI shot, the sequence blips through the phase encoding dimension (P), while varying prephasers and rephasers in the slice encoding dimension (S). In this way, one k-space plane per shot is acquired with repetition time $T_R$, yielding a full volume after $N_z$ slice encodings with volume acquisition time $T_{vol}=T_R \cdot N_z$.

Servo navigation \cite{ulrich_servo_2024,riedel_run-time_2025} relies on three basic components (Fig. \ref{fig:fig1_sequence_control}B): a navigator, a model for parameter estimation, and a mechanism for run-time scan control. First, a 3D orbital navigator, acquiring orbits around three orthogonal axes sequentially, at a k-space radius of $400\unit{rad/m}$, is inserted between excitation and the echo-planar readout as shown in Fig. \ref{fig:fig1_sequence_control}A. The duration of the navigator $\tnav$ can be varied and is, by default, set to $3.2\unit{ms}$. Second, a linear model of signal change as a function of motion, field, and phase parameters is determined on the fly, from four reference navigators taken in a fraction of a second \cite{ulrich_servo_2024}. The model matrix $M$ \cite{ulrich_servo_2024} relates the parameter vector $\vec{\Delta}$ to change in navigator signal: 
\begin{equation} 
	\vec{s}\left(\vec{\Delta}\right)-{\vec{s}}^{\,(0)}=M\vec{\Delta},
	\label{eq:linproblem}
\end{equation}
where the vectors $\vec{s}\left( \vec{\Delta} \right)$ and ${\vec{s}}^{\, (0)}$ contain the complex-valued signals of the current navigator and a reference navigator and $\vec{\Delta}$ includes six 3D rigid motion parameters, a global phase $\Delta\phi$ and a global frequency parameter $\Delta\omega$. ${\vec{s}}^{\, (0)}$ is acquired as a separate (fifth) reference navigator to decorrelate noise between the matrix columns and the navigators \cite{riedel_run-time_2025}. Third, run-time updates counter motion and field changes by rotating gradients and shifting slab-selective excitation. This process forms a control loop that stabilizes both the imaging process and the navigator acquisition itself, keeping the signal model valid beyond the initial linear range.

\hypertarget{methods-navi-f0-correction}{%
	\subsection{Navigator-based phase correction}\label{methods-navi-f0-correction}}
Relative parameter updates of global phase $\Delta\phi$ and frequency $\Delta\omega$ are accumulated over the scan as described in Ref. \cite{ulrich_servo_2024}, yielding absolute time courses for the total global phase $\Delta\phi_{total,j}$ and frequency $\Delta\omega_{total,j}$, with shot index $j$. Phase correction of image raw data is then achieved by the phase shift:
\begin{equation}
    \Delta\varphi_j(t)=\Delta\omega_{corr,j}\,t+\Delta\phi_{corr,j},
    \label{eq:nav_phase_corr_abs}
\end{equation}
where $t$ is the intra-shot sampling time. $\Delta\omega_{corr,j}$ and  $\Delta\phi_{corr,j}$ are the navigator traces for the total global phase and frequency after optional filtering over time:
\begin{align}
    \Delta\phi_{corr,j}&=\mathrm{filter}_j(\Delta\phi_{total,j}),\label{eq:nav_filtering1}\\
    \Delta\omega_{corr,j}&=\mathrm{filter}_j(\Delta\omega_{total,j}).
    \label{eq:nav_filtering2}
\end{align}
The phase correction in Eq. \ref{eq:nav_phase_corr_abs} is, however, prone to noise amplification for long echo times or sampling durations, because the noise in the frequency parameter $\Delta\omega_{corr,j}$ is amplified linearly with the acquisition time $t$. Requirements for field parameter precisions, therefore, depend on the sequence timing. Filtering (Eqs. [3-4]) serves to reduce noise amplification, at the expense of effective correction bandwidth. Alternatively, the navigator acquisition time $\tnav$ can be made longer to increase field sensitivity and SNR at the expense of time and flexibility.

As an alternative strategy, phase correction with relative timing can be used to suppress noise amplification for short volume acquisitions:
\begin{equation}
    \Delta\varphi_{\mathrm{rel},j}(t)=\Delta\varphi_j(t-T_E)=\Delta\omega_{corr,j}\,\cdot\,(t-T_E)+\Delta\phi_{corr,j}.
    \label{eq:nav_phase_corr_rel}
\end{equation}
Here, phase correction is performed with timing $t_{rel}=(t-T_E)$ relative to echo time $T_E$ and neglects the shot-wise phase $\Delta\omega_{corr,j}\,T_E$ within a volume acquisition. By this, the phase along the k-space line $(0,0,k_z)^T$ remains unchanged. Relative timing is beneficial when the frequency variation of $\Delta\omega_{corr,j}$ within a volume acquisition is dominated by noise rather than actual field evolution, which in our experience was typically the case for fast fMRI volume acquisitions of a few seconds at 3T.

\hypertarget{methods-epi-f0-correction}{%
	\subsection{PEERS for EPI trajectories}\label{methods-epi-f0-correction}}
fMRI scans possess a strongly repetitive structure with minor local BOLD-induced contrast variations. The redundancy along the EPI time series can be leveraged to perform self-navigated frequency and phase corrections per shot relative to a peer shot from the reference volume as proposed by Parker et al. \cite{parker_k-space-based_2022}. Head rotations, however, violate the k-space correspondences between the shot peers and can potentially render the method ineffective or even harmful. We propose phase equalization enhanced by run-time stabilization (PEERS), combining the phase equalization by Parker et al. \cite{parker_k-space-based_2022} with run-time motion correction.

\begin{figure}[H]
    \centering
	\ifincludeplots\optincludegraphics[width=\textwidth]{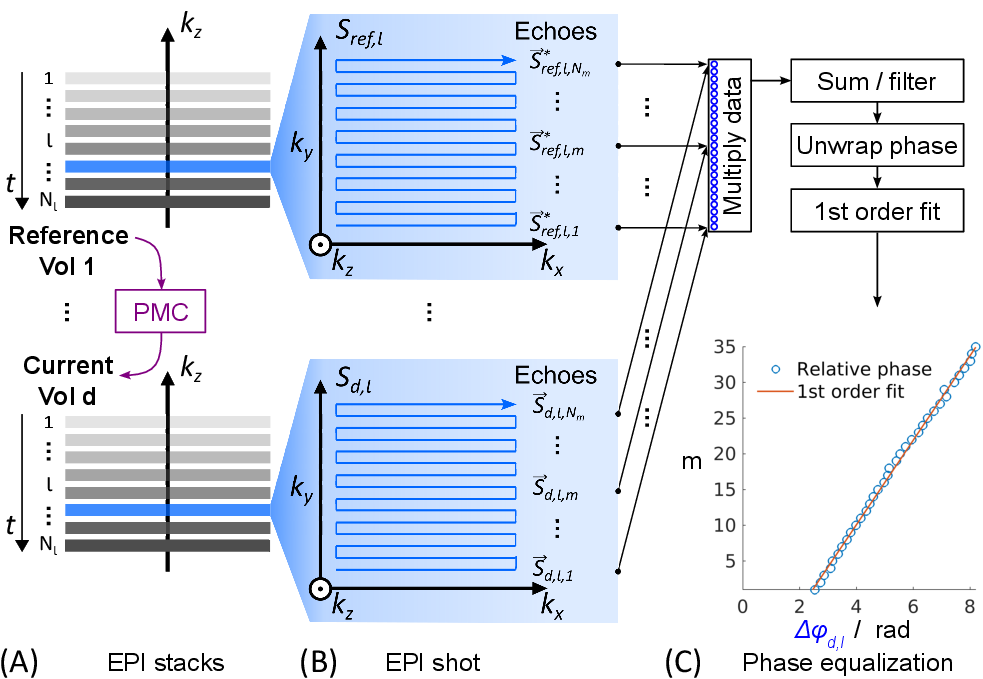}\fi
	\caption{Phase equalization enhanced by run-time stabilization (PEERS). (A) Shots from the first EPI stack (volume) serve as a reference for phase equalization of subsequent shots after prospective motion correction (PMC). (B) For any subsequent shot from volume $d$, relative phases are calculated echo-by-echo by scalar products of each echo signal $\vec{S}_{d,l,m}$ with its respective reference peer. $m$ and $l$ are echo and slice encoding indices. (C) Phases are unwrapped along the EPI echo train ($m$) followed by a first-order fit yielding frequency (slope) and phase (offset) parameters.}
	\label{fig:fig2_epicorrOverview}
\end{figure}

Algorithm \ref{alg:phase_equalization} describes the retrospective processing similar to Ref. \cite{parker_k-space-based_2022}.
Each volume repetition with index $d$ consists of $N_l$ EPI segments with a given trajectory, and each segment $l$ is composed of $N_m$ echoes, forming the echo train. Each repetition-segment pair ($d$, $l$) is associated with a running shot index $j=l+(d-1)N_l$ over the whole scan. An EPI shot obtains a 3D data array $S_{d,l}^{\,}$ of shape $N_m{\times}N_k{\times}N_c$, where $m$, $k$, and $c$ are the echo, readout sample, and coil indices. $T_{l}$ is a $N_m{\times}N_k{\times}1$ array comprising the EPI readout timing. A shot-wise global phase $\Delta\phi_{\subPeers,j}$ and frequency $\Delta\omega_{\subPeers,j}$ is estimated for each shot from the phase difference vector $\Delta\vec\varphi_{d,l}^{\,}$ ($N_m\times1$) to its peer. An overview of the PEERS method is shown in Fig. \ref{fig:fig2_epicorrOverview}. The method without run-time motion correction is simply termed \emph{phase equalization}.

\begin{algorithm}
\caption{Self-navigated phase equalization for segmented 3D EPI based on Ref. \cite{parker_k-space-based_2022}. $\circ$, $^*$, and $\measuredangle$ are element-wise product, complex conjugation and phase extraction. 
}\label{alg:phase_equalization}
\begin{algorithmic}
\For{$j = 1$ to $N_j$}
    \State $d \gets \mathrm{ceil}\left(j\,/\,N_l\right), \quad l \gets j \bmod{N_l}$
    \If{$d=1$} \Comment{First volume repetition}
        \State $S_{\subRef,l}^{\,}=S_{d,l}^{\,}$ \Comment{Set reference shot (peer)}
    \Else
        \State $\vec{p}_{d,l}^{\,}\gets\mathrm{sum}_{k,c}^{\,}\left(S^{^{\,*}}_{\subRef,l}\;{\circ}\;S_{d,l}^{\,}\right)$ \Comment{Scalar product over readout \& coils}
        \State $\Delta\vec\varphi_{d,l}^{\,}\gets\mathrm{unwrapPhase}_m^{\,}\left(\measuredangle\;\vec{p}_{d,l}^{\,}\right)$ \Comment{Unwrap phase along echo train ($m$)}
        \State $\left[\Delta\omega_{\subPeers,j}^{\,}, \Delta\phi_{\subPeers,j}^{\,}\right]\gets\mathrm{linearFit}_m^{\,}\left(\Delta\vec\varphi_{d,l}^{\,}\right)$ \Comment{First-order fit along echo train ($m$)}
        \State $S_{d,l}^{\,}\;{\gets}\;S_{d,l}^{\,}\;\circ\;\exp\left(-i\left(\Delta\omega_{\subPeers,j}^{\,}\,T_{l}+\Delta\phi_{\subPeers,j}^{\,}\right)\right)$ \Comment{EPI phase correction}
    \EndIf
\EndFor
\end{algorithmic}
\end{algorithm}

This algorithm corrects shot phase variations to the first temporal order with respect to its peer and, by this, matches the frequency- and phase-induced point spread function to that of the reference volume, leading to stabilized voxel time series. Unaware of the causes of the phase variations, the correction includes both system- and physiology-related effects like field changes, RF phase drifts, and breathing \cite{parker_k-space-based_2022}. The sensitivity of the EPI sampling to frequency variations naturally adapts to variations in echo time, sampling duration, and field strength. The technique can generally handle arbitrary EPI phase encoding schemes as long as they are repeated for each volume.

\hypertarget{methods-arbitrary-f0-correction}{%
	\subsection{Phase equalization for arbitrary trajectories}\label{methods-arbitrary-f0-correction}}
Phase equalization based on Ref. \cite{parker_k-space-based_2022} is tailored to EPI trajectories with separable echoes. An extension of Algorithm \ref{alg:phase_equalization} to arbitrary trajectories is possible by collapsing the echo ($m$) and intra-echo readout ($k$) dimensions of the EPI signal array $S_{d,l}^{\,}$ ($N_m{\times}N_k{\times}N_c$) into a collapsed readout ($k^\prime$) dimension ($N_{k^\prime}{\times}N_c$). The EPI-specific implementation uses the noise-robust complex-valued sum over the echo-wise readout dimension, which represents an inherent magnitude-weighting for the phase estimation as described in Ref. \cite{parker_k-space-based_2022}. For arbitrary trajectories, the complex-valued sum can be replaced by mean filtering along the collapsed readout dimension ($k^\prime$) sorted in time, yielding the complex-valued vector $\vec{p}_{d,l}^{\,}\gets\mathrm{meanFilter}_{k^\prime}^{\,}\left(\mathrm{sum}_{c}^{\,}\left(S^{^{\,*}}_{\subRef,l}\;{\circ}\;S_{d,l}^{\,}\right)\right)$ of length $N_{k^\prime}$. Finally, the time array $T_l$ needs to be adapted to the collapsed readout dimension with shape $N_{k^\prime}\times1$ for the final phase correction.
    
\hypertarget{methods-AQ-details}{%
	\subsection{Acquisition details}\label{methods-aq-details}}
The scan protocol consisted of three 3D EPI runs lasting $10\unit{min}$ each: (i) without servo navigation, (ii) with servo navigation ($\tnav=3.2\unit{ms}$), and (iii) with servo navigation and a long navigator ($\tnav=11.5\unit{ms}$). For case (iii), the navigator trajectory was stretched in time while keeping the same number of samples to increase the SNR. The order of the three runs was randomized and unknown to the subject. The scan parameters were: $2.5\unit{mm}$ isotropic resolution, FOV=$230\times200\times140\unit{mm^3}$ (incl. $1.4$-fold slice oversampling), $N_z=32$ slice encoding steps with linear ordering, reduction factors $2.2 \times 1.75$ (phase $\times$ slice), flip angle=$17.2\unit{deg}$, TE=$30\unit{ms}$, $T_R$=$64\unit{ms}$, $T_{vol}$=$2.0\unit{s}$, and 300 volumes after 10 initial dummy volumes to build up the steady state. SPIR fat suppression \cite{kaldoudi_chemical_1993} was used and its spoiler gradients were kept static in the initial orientation, because occasionally servo navigation instabilities were observed that probably stem from mechanical vibrations. A T1-weighted MP-RAGE scan \cite{mugler_iii_rapid_1991} was acquired at isotropic resolution $1.1\unit{mm}$ for anatomical reference.

The method was evaluated in a phantom and in-vivo. For the phantom experiments, a sphere filled with small PMMA balls and a copper sulfate solution was used as described in Ref. \cite{ulrich_servo_2024}. In-vivo, the scan sets were performed in six healthy volunteers in accordance with local ethical regulations on a 3T Philips (Best, The Netherlands) Ingenia scanner with a 32-channel head coil. A visuomotor fMRI task was performed with a fully randomized event-related design showing images of arrows presented in the upper or lower visual field. Subjects were asked to perform left- or right-handed button presses according to the direction of the displayed arrows. Respiratory and cardiac physiology was monitored using a breathing belt and a pulse oximeter.

To study the influence of motion, 3-min scans (100 volumes) with and without servo navigation were added for two of the six subjects with instructed motion. Subjects repeated a succession of rotations starting from a reference position and changing pose every $10\unit{sec}$ to [$0$, $+1^\circ (p)$, $0$, $-1^\circ (p)$, $0$, $+1^\circ (y)$, $0$, $-1^\circ (y)$] with $p$ and $y$ being pitch (nod) and yaw (head shake) rotations.

\hypertarget{methods-REC-details}{%
	\subsection{Reconstruction details}\label{methods-rec-details}}
The image reconstruction was implemented in MATLAB R2022b (MathWorks Inc., Natick, MA, USA) using the ReconFrame pipeline (Gyrotools, Zurich, Switzerland) with several modifications. After some basic vendor-specific corrections, the k-space data was corrected for the total global phase and frequency parameters of the navigators' run-time estimates with relative timing following the descriptions in Section \ref{methods-navi-f0-correction}. A non-causal median filter of size 9 ($\sim 0.6\unit{s}$) was used for parameter filtering. After gridding of the EPI ramp samples, the k-space data was corrected for the shifts in measurement and phase direction according to the navigators' run-time shift estimates. This was done in the reconstruction because only the slab shift was treated in run-time by the scanner.

Next, EPI-based phase equalization was performed estimating shot-wise phases and frequencies by self-navigation as described in Section \ref{methods-epi-f0-correction}. A Nyquist ghosting correction was calculated from a vendor-provided pre-scan and applied in hybrid space followed by a ReconFrame implementation of regularized SENSE \cite{pruessmann_sense_1999} in image space. 
All navigator-based corrections were turned off if servo navigation was disabled.
Sensitivity maps were reconstructed from the vendor's reference scan in the beginning of the scan set and included masking and smoothing. 

Statistical analysis was performed using SPM12 \cite{penny_statistical_2011}, including volume realignment \cite{friston_movement-related_1996}, structural image co-registration, and tissue segmentation into gray and white matter, CSF, skull, soft tissue, and background \cite{ashburner_unified_2005}. Physiological traces of the respiration belt and the pulse oximeter were synchronized with the fMRI data using PhysIO \cite{kasper_physio_2017}. tSNR was calculated before and after SPM volume realignment as the voxel-wise mean over time divided by the voxel-wise standard deviation (std) over time: $\mathrm{tSNR}_{\boldsymbol{r}}={\mathrm{mean}_\nu(\rho_{\boldsymbol{r},\nu})} / {\mathrm{std}_\nu(\rho_{\boldsymbol{r},\nu})}$,
where $\rho_{\boldsymbol{r},\nu}$ is the magnitude signal of a voxel at location $\boldsymbol{r}$ and volume (time) index $\nu$.


\hypertarget{results}{%
	\section{Results}\label{results}}

\begin{figure}[H]
    \centering
	\ifincludeplots\optincludegraphics[width=\textwidth]{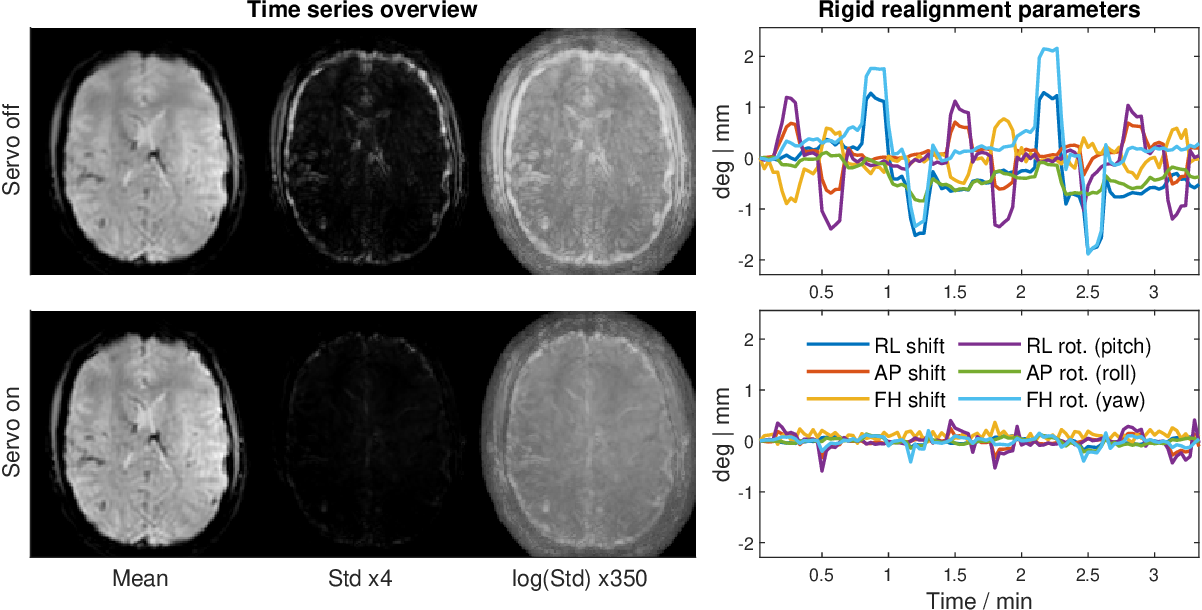}\fi
	\caption{In-vivo example with instructed motion comparing \pmcOff\, and \pmcOn. (Left): Mean and standard deviation (Std) of the voxel time series. (Right): realignment parameters. For \pmcOff, the mean image is blurred and the std shows strong edge enhancement and choppy patterns in the logarithmic standard deviation, which are clearly mitigated by \pmcOn. The realignment parameters for \pmcOff\, show the instructed motion pattern, while being close to zero for \pmcOn\, due to effective run-time motion control.}
	\label{fig:fig3_realignmentExampleInvivoMotion}
\end{figure}

\hypertarget{results-moco-example}{%
	\subsection{Run-time motion correction example}\label{results-moco-example}}
Figure \ref{fig:fig3_realignmentExampleInvivoMotion} shows an in-vivo example with instructed motion for \pmcOff\, and \pmcOn\, comparing mean and standard deviation (std) of the voxel time series as well as the realignment parameters. For \pmcOff, the central slice of the mean volume appears blurred, while the standard deviation shows strongly enhanced edges and choppy patterns throughout the brain in the logarithmic image. In contrast, the blurring is mitigated by \pmcOn, and the imprint of anatomical features and choppy patterns in the standard deviation images is clearly reduced. Edge enhancement mainly originates from frequency drift-induced EPI shifting that is corrected by servo navigation, whereas choppy patterns are caused by motion-induced inconsistencies.

The volume realignment parameters for \pmcOff\, show the instructed motion pattern of successive pitch and yaw rotations, which are left uncorrected in the data. For \pmcOn, the volume realignment detects only minor motions close to zero that are left after the run-time motion correction. However, the onset of the motion events is still visible with \pmcOn\, by small spikes and plateaus, especially for the pitch rotation (purple).

\begin{figure}[H]
    \centering
	\ifincludeplots\optincludegraphics[width=0.6\textwidth]{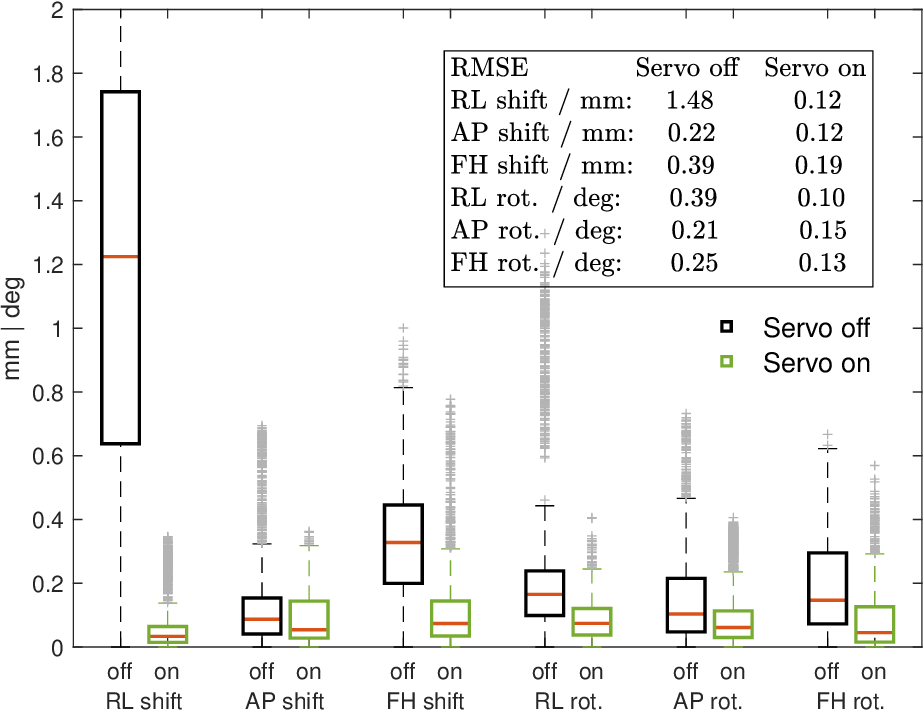}\fi
	\caption{Volume realignment statistics for all in-vivo subjects without instructed motion comparing \pmcOff\, (black boxes) and \pmcOn\, (green boxes). Boxplots for shifts along RL, AP, and FH are shown, as well as the respective rotations. A table of parameter RMSE values is included. Motion is consistently reduced by servo navigation for all parameters. The RL shift (phase encoding direction), which includes both subject motion and EPI shifting from frequency drift, shows the strongest reduction.}
	\label{fig:fig4_realignmentStatisticsInvivo}
\end{figure}

\hypertarget{results-realignment-stats}{%
	\subsection{Realignment parameter statistics}\label{results-realignment-stats}}
Figure \ref{fig:fig4_realignmentStatisticsInvivo} shows the volume realignment parameter statistics and RMSE results for the data from all subjects without instructed motion for \pmcOff\, (black boxes) and \pmcOn\, (green boxes). RMSE, median realignment parameters, and inter-quartile ranges are consistently reduced by servo navigation for all parameters. The RL shift includes not only involuntary shifts of the subjects but also the EPI shifting in phase encoding direction due to frequency drifts as the volume realignment is incapable of differentiating between the two. As the navigator corrects both for subject- and frequency-induced shifts in phase encoding direction, the RL shift is most prominently reduced.

\begin{figure}[H]
    \centering
	\ifincludeplots\optincludegraphics[width=\textwidth]{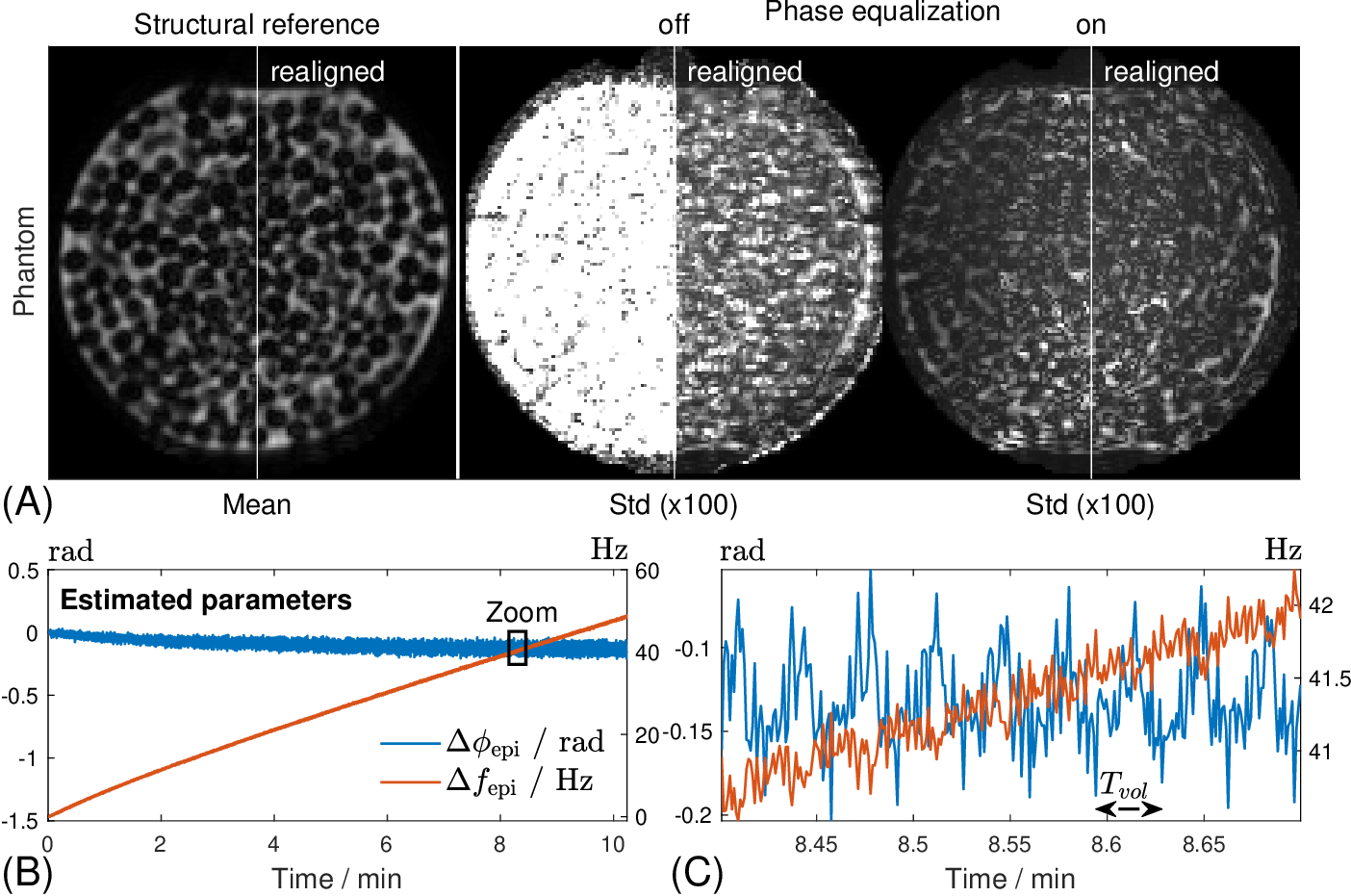}\fi
	\caption{Phase equalization of voxel times series in a phantom scan without servo navigation. (A): Mean and standard deviations (Std) without and with phase equalization. Each image is split by half showing the image before and after realignment. (B): Phase $\Delta\phi_{\subPeers}$ and frequency ${\Delta}f_{\subPeers}=\Delta\omega_{\subPeers}/(2\pi)$ estimated by phase equalization. (C): Zoom into (B). EPI shifting from the frequency drift is reduced by realignment without phase equalization. Phase equalization captures the frequency drift well (B) and further improves volume consistency by intra-volume phase correction. Repetitive patterns with $T_{vol}$-period are visible in (C).}
	\label{fig:fig5_shotPhaseEqualizationExample}
\end{figure}

\hypertarget{results-peers-example}{%
	\subsection{Phase equalization example}\label{results-peers-example}}
Figure \ref{fig:fig5_shotPhaseEqualizationExample} shows phase equalization in the phantom without servo navigation. The mean of the voxel time series is shown for reference. Without phase equalization, the standard deviation (std) is the highest and is strongly reduced by realignment. With phase equalization, the standard deviation is further reduced, and realignment does not give an additional advantage. A frequency drift of about $5\unit{Hz}/\unit{min}$ is captured in the estimated parameters (B), and the zoom (C) shows patterns that repeat with volume acquisition time $T_{vol}$.

Continuous frequency drift is a major cause of tSNR instability, which appears as EPI shifting in the low-bandwidth phase encoding direction \cite{kasper_monitoring_2015}. Volume realignment is well suited to correct for these inter-volume shifts, strongly reducing the standard deviation for \emph{phase equalization off}. While realignment can only correct for global transformations between volumes, the phase equalization is capable of addressing both inter- and intra-volume phase and frequency variations including slow frequency drifts and sampling-related effects. By this, phase equalization reduces the respective artifacts and thus improves volume consistency.

\begin{figure}[H]
    \centering
	\ifincludeplots\optincludegraphics[width=\textwidth]{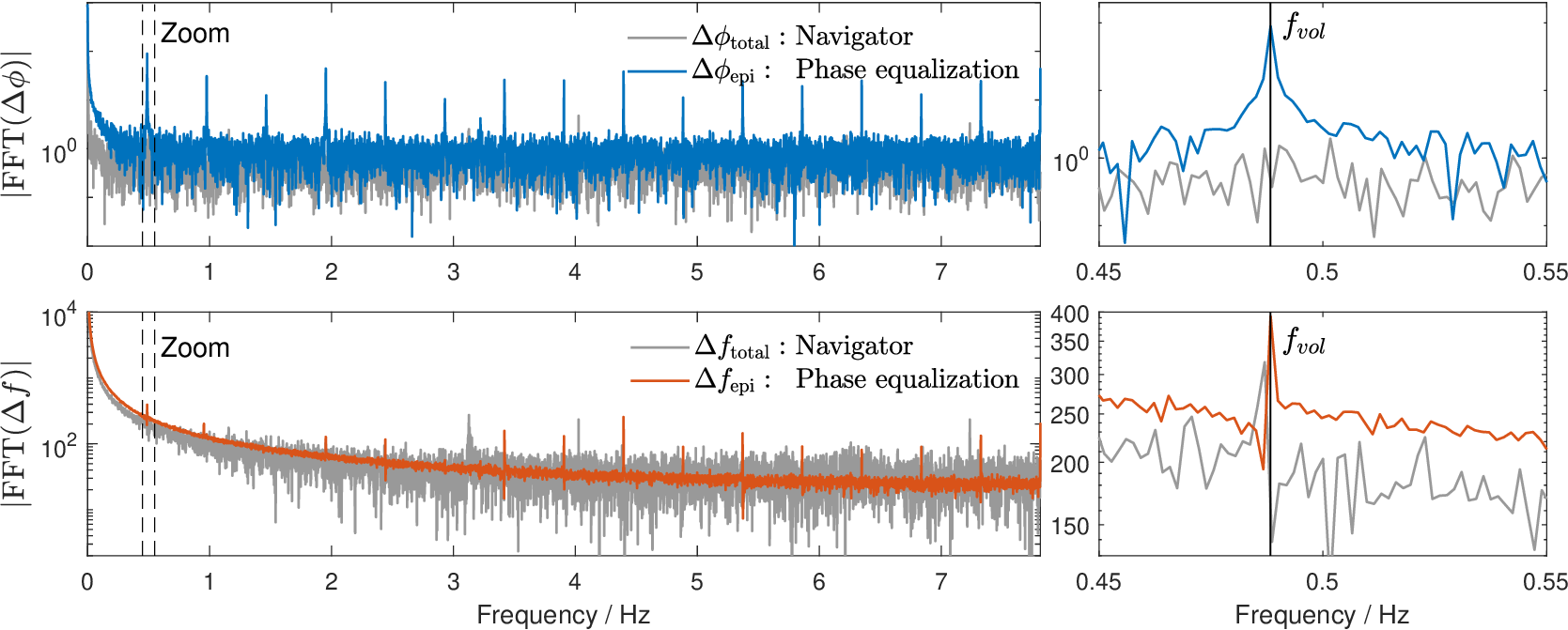}\fi
	\caption{Phase equalization parameter spectra for phantom scan without servo navigation in Figure \ref{fig:fig5_shotPhaseEqualizationExample}. (Top row): Phase power spectrum. (Bottom row): Frequency power spectrum. Navigator parameter spectra are shown for reference. Zooms on the volume frequency $f_{vol}$ are provided on the right. While spectral contents at $f_{vol}$ are visible in both datasets, the amplitude and Dirac-comb pattern indicates more prominent slice encoding-related gradient characteristics in the EPI signal, which is corrected for by self-navigated phase equalization.}
	\label{fig:fig6_shotPhaseEqualizationSpectra}
\end{figure}

\hypertarget{results-peers-spectra}{%
	\subsection{Spectral parameter analysis}\label{results-peers-spectra}}
Figure \ref{fig:fig6_shotPhaseEqualizationSpectra} shows the power spectra of the phase equalization parameters compared to the respective navigator spectra. The phase equalization parameters contain a Dirac-comb pattern spaced by $f_{vol}=1/T_{vol}\approx0.49\unit{Hz}$, which is associated with the repetitive $T_{vol}$-spaced patterns of the time domain signals in Fig. \ref{fig:fig5_shotPhaseEqualizationExample}C. The navigator frequency estimates also show a peak at $f_{vol}$ that is slightly smaller. The patterns thus match the volume encoding frequency, which indicates that the EPI data is affected by transient field effects that differ from those acting on the navigator and are therefore not correctable by the navigator.

\begin{figure}[H]
    \centering
	\ifincludeplots\optincludegraphics[width=\textwidth]{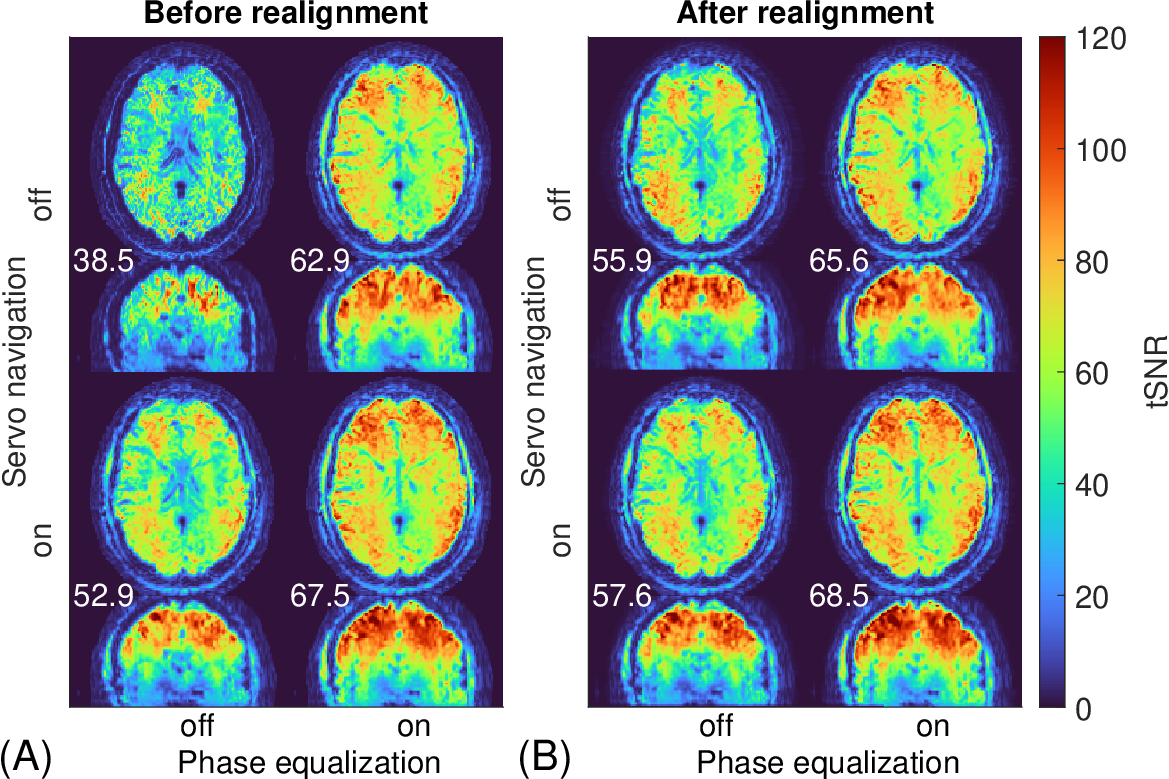}\fi
	\caption{In-vivo tSNR example of one subject without instructed motion comparing the performance of servo navigation and phase equalization. (A) Before realignment. (B) After realignment. Mean tSNR values for whole brain are given in the images. Without any of the three corrections (top left) tSNR is lowest mainly due to the frequency drift. Both servo navigation and phase equalization improve tSNR individually, but are most effective together in PEERS. Realignment further contributes a small tSNR improvement.}
	\label{fig:fig7_tsnrExampleInvivo}
\end{figure}

\hypertarget{results-tsnr-example}{%
	\subsection{In-vivo tSNR example}\label{results-tsnr-example}}
Figure \ref{fig:fig7_tsnrExampleInvivo} compares the tSNR in-vivo with and without servo navigation, phase equalization and volume realignment. The worst tSNR performance appears when all options are disabled. By enabling the individual corrections one by one, tSNR improves consistently, showing the best result with all corrections enabled. If both servo navigation and phase equalization are enabled, volume realignment can only slightly improve the results.
Note that \pmcOff\, and \pmcOn\, are two separate datasets and could be affected by different underlying motions, while phase equalization, as a pure reconstruction technique, is turned off and on for the exact same raw data.

\hypertarget{results-tsnr-stats}{%
	\subsection{In-vivo tSNR statistics}\label{results-tsnr-stats}}
Table \ref{tab:tsnrStats_shortNav_latex} shows the mean tSNR over all subjects for different tissue types after realignment including gray matter (GM), white matter (WM), cerebrospinal fluid (CSF), and their combination (Whole brain). \emph{Whole brain - not aligned} is given as a reference excluding realignment. 

Both techniques consistently increase tSNR for all anatomies. Without realignment, phase equalization only shows minor improvements in tSNR ($8\%$) compared to $45\%$ with servo navigation. Combining both servo navigation and phase equalization, PEERS achieves an overall whole brain tSNR increase of $60\%$. The main tSNR degradation factor that is corrected here is the frequency drift. With volume realignment, the improvements in the whole brain are slightly greater for phase equalization ($6\%$) than for servo navigation ($3\%$). Together, an increase in tSNR of $12\%$ is achieved, which is more than adding the individual contributions of servo navigation and phase equalization, supporting the synergistic relationship. Furthermore, the tSNR increase after volume realignment quantifies the positive impact of intra-volume motion and field corrections, which are not achievable by inter-volume corrections in the post-processing. Note that the relative impact of motion on tSNR depends on the prevalence of motion in the data and these scans were performed without motion instructions and included only healthy, cooperative volunteers.
    
\begin{figure}[H]
    \centering
	\ifincludeplots\optincludegraphics[width=\textwidth]{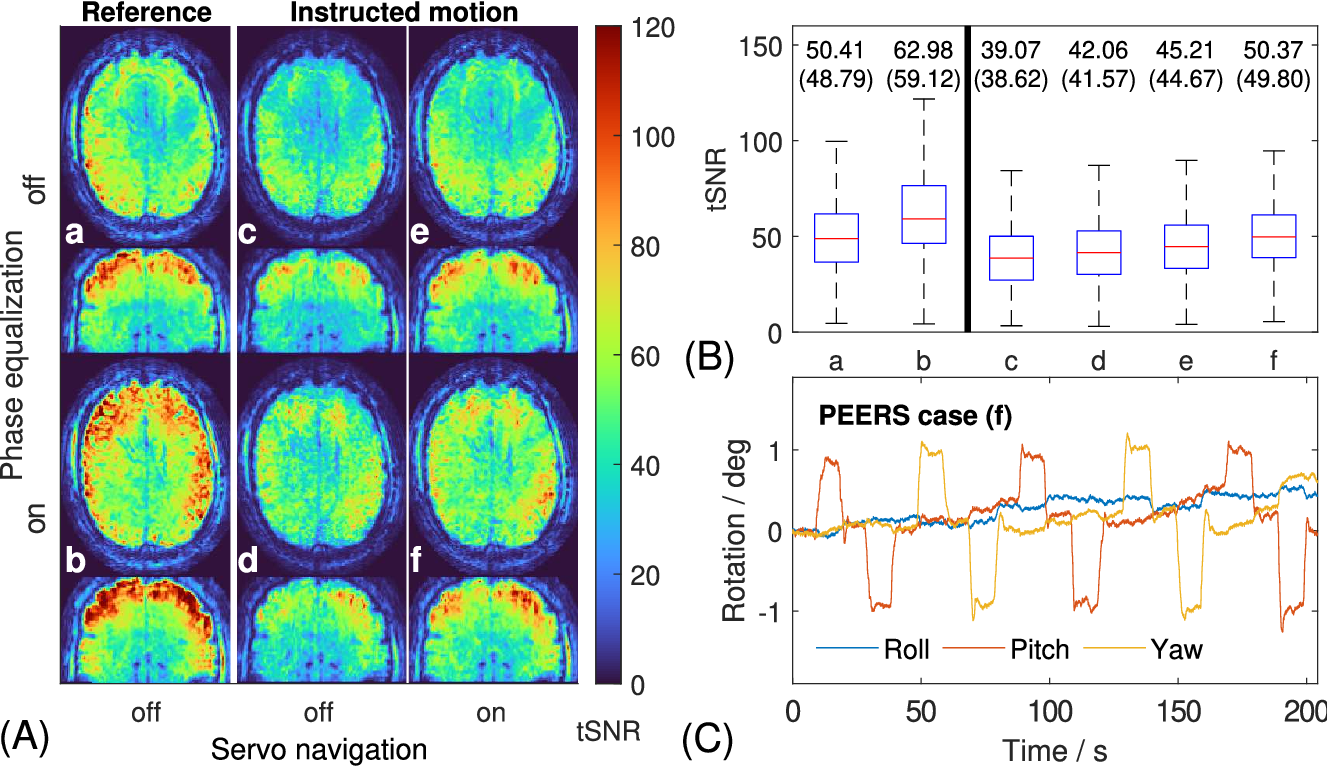}\fi
	\caption{In-vivo tSNR example with instructed motion comparing the performance of servo navigation and phase equalization after realignment. (A) tSNR images without (a-b) and with (c-f) motion. (B) Associated tSNR boxplots with mean (median) values. (C) Navigator-based rotation parameters for \pmcOn. The boxplots show a clear tSNR reduction by motion without servo navigation and phase equalization (c). With motion, the servo navigation contributes higher tSNR gains than phase equalization, although both consistently improve tSNR.}
	\label{fig:fig8_tsnrMotionInvivo_Subject2}
\end{figure}

\hypertarget{results-synergy}{%
	\subsection{PEERS in the presence of motion}\label{results-synergy}}
Figure \ref{fig:fig8_tsnrMotionInvivo_Subject2} shows a tSNR comparison for an in-vivo case with instructed motion after realignment. With motion, the mean tSNR drops from about 50 (a; reference) to 39 (c). The mean tSNR increases to 42 for phase equalization alone (d) and to 45 for servo navigation alone (e), while reaching 50 for PEERS, which includes both corrections (f). Still, the corrected motion case does not achieve the tSNR performance of 63 of the motion-free case with phase equalization. If motion is present, the relative importance of the run-time motion correction thus increases. The in-vivo results are supported by phantom data shown in Supporting Figure \ref{fig:figS1_tsnrMotionPhantom}, where the phantom was placed on a platform and shifted in and out by $1\unit{mm}$ every 10 seconds. The shifting direction was aligned with the frequency, phase and slice direction in separate scans.

\begin{figure}[H]
    \centering
	\ifincludeplots\optincludegraphics[width=\textwidth]{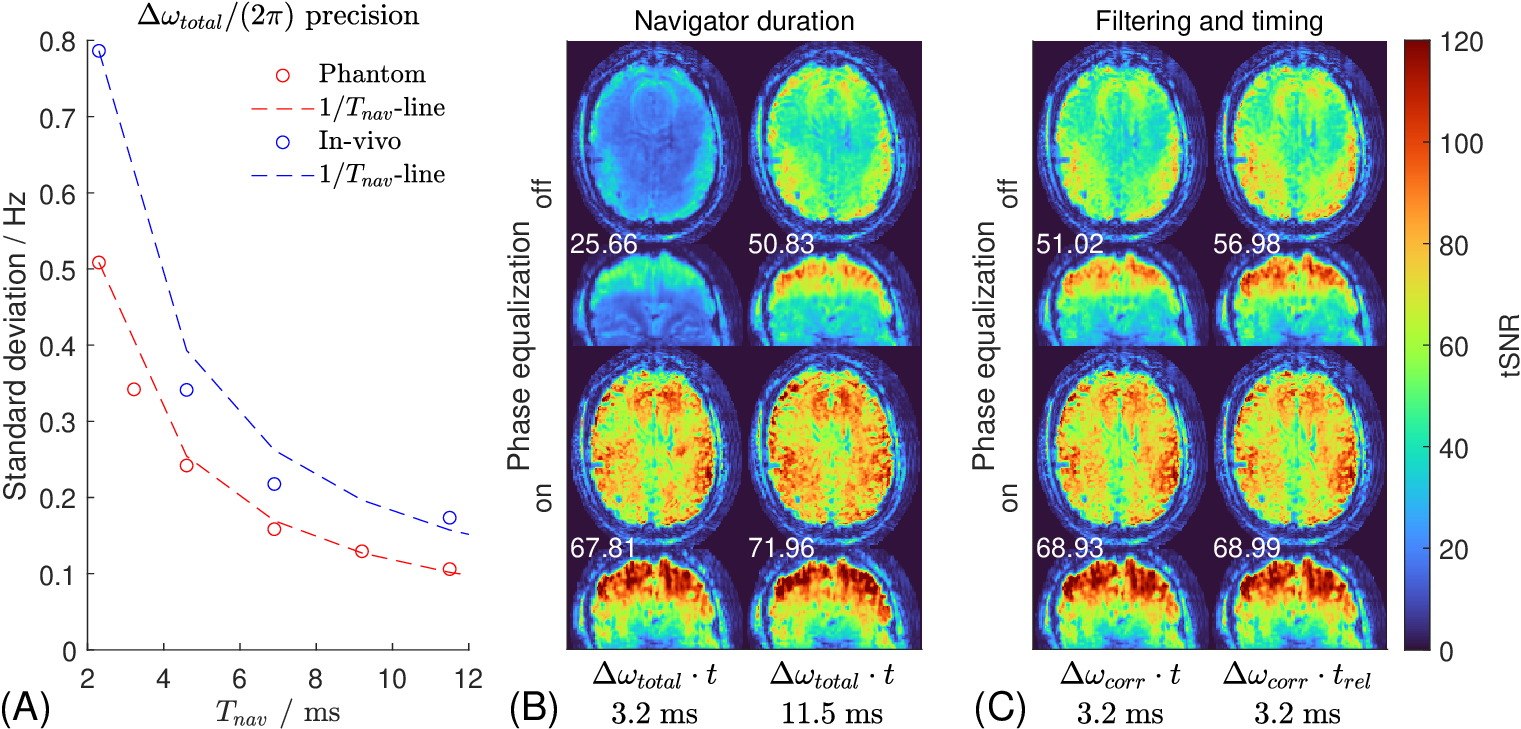}\fi
	\caption{Considerations regarding navigator- and EPI-based frequency estimates for image correction. (A) Frequency precision for different navigator durations $T_{nav}$. (B) tSNR impact of phase equalization and $T_{nav}$. (C) tSNR impact of filtering and relative timing $t_{rel}$. The phase correction type and $T_{nav}$ are indicated below the images. Phase correction is either without ($\Delta\omega_{total}$) or with ($\Delta\omega_{corr}$) filtering and with absolute ($t$) or relative ($t_{rel}$) timing. Long navigators, filtering and relative timing all improve navigator frequency corrections. EPI-based phase equalization performs better but requires a repetitive sequence structure.}
	\label{fig:fig9_longNavigatorExample_rel}
\end{figure}

\hypertarget{results-long-navi}{%
	\subsection{Frequency precision considerations}\label{results-long-navi}}
Figure \ref{fig:fig9_longNavigatorExample_rel} characterizes the navigator's frequency precision, its impact on tSNR and various ways to mitigate noise propagation into the phase correction including also PEERS. In (A), the standard deviation of the (unfiltered) frequency $\Delta\omega_{total}$ is plotted for different navigator durations $T_{nav}$ for phantom and in-vivo scans. The frequency precision grows approximately linearly with the duration of sampling. $1/T_{nav}$-guidance lines starting from the first $2.3\unit{ms}$-navigator value are given to visualize the trend. tSNR maps for navigator durations of $3.2\unit{ms}$ and $11.5\unit{ms}$ with and without PEERS are shown in Fig. \ref{fig:fig9_longNavigatorExample_rel}B. Phase corrections with unfiltered frequency $\Delta\omega_{total}$ from short navigators result in poor tSNR, which is improved by longer navigators due to increased precision. As shown in (C), phase corrections with filtered frequencies $\Delta\omega_{corr}$ (Eqs. \ref{eq:nav_filtering1}-\ref{eq:nav_filtering2}) and with relative timing $t_{rel}$ (Eq. \ref{eq:nav_phase_corr_rel}) considerably improve tSNR also for short $3.2\unit{ms}$-navigators. Nevertheless, EPI-based PEERS corrections, which rely on the repetitive sequence structure, outperform navigator corrections in all cases.



\hypertarget{discussion}{%
	\section{Discussion}\label{discussion}}
In this work, servo navigation has been successfully integrated into a segmented 3D EPI sequence for run-time motion corrected fMRI. The method offers on-the-fly self-calibration without additional hardware or extra steps in the workflow. Various methods to reduce noise propagation from the navigator's frequency estimates into the EPI phase corrections have been investigated to enable servo navigation with navigator-based phase correction. Enabled by run-time motion correction, PEERS offers robust shot-wise phase and frequency corrections by exploiting redundancies along EPI time series. Servo navigation mitigates artifacts orignating from both intra- and inter-volume motion, and, in combination with PEERS, improves the stability of voxel time series beyond rigid realignment. The stabilizing phase corrections of PEERS enable servo navigation with short $3.2\unit{ms}$-navigators even for fMRI sequences with long echo times, underlining the synergistic relationship of PEERS.

\hypertarget{discussion-servo-navigation}{%
	\subsection{Servo navigation for 3D EPI}\label{discussion-servo-navigation}}
In Figs. \ref{fig:fig3_realignmentExampleInvivoMotion}-\ref{fig:fig4_realignmentStatisticsInvivo}, servo navigation has been shown to capture motion and frequency drifts well, clearly reducing the amount of motion in the imaging raw data. However, residual motion has been observed even with servo navigation in the form of remaining spikes at the onset of instructed pitch rotations and small plateaus while the position was held. The onset spikes are probably caused by the latency of the control system \cite{zaitsev_magnetic_2006}. Our servo navigation implementation currently requires about $300\unit{ms}$ to apply an estimated motion state. Small plateaus in pitch-rotated positions could be caused by secondary motion effects such as pose-induced coil sensitivity or field changes \cite{maclaren_prospective_2013}, which can induce parameter biases \cite{riedel_run-time_2025}. Navigator signal variations from coil sensitivity changes should generally be present for both instructed rotations, while only nodding rotations are related to susceptibility-induced field changes \cite{marques_application_2005}.
As the plateaus are mainly seen for nodding, the biases are probably related to pose-dependent field changes.

It is important to note that the volume realignment, which serves as a reference for motion quantification in this study, is also biased by pose-dependent contrast variations, such as varying distortions or signal dropouts around the nasal cavities after pose changes.
Retrospective volume realignment itself should benefit from run-time motion correction, because the registered volumes are more consistent, and updating the phase encoding direction in run-time reduces geometric distortions.

\hypertarget{discussion-peers}{%
	\subsection{Phase equalization}\label{discussion-peers}}
Phase equalization has been shown to detect and correct frequency drifts and slice encoding-related sampling effects in the phantom (Figs. \ref{fig:fig5_shotPhaseEqualizationExample}-\ref{fig:fig6_shotPhaseEqualizationSpectra}). The latter are probably caused by long-term eddy currents that are generated by the slice encoding pre-phasers directly prior to the EPI readout.
Using a first-order phase fit over the echo train, the model can handle only field variations constant over the echo train as caused by long-term eddy currents. High-frequency variations, such as those arising from vibrations or short-term eddy currents, cannot be modeled in this way but were not prominent in the present data.

The stability improvements by phase equalization have been reproduced on two different scanners of the same vendor in our lab. As sampling-related parameter variations are related to general gradient imperfections, the relevance of this correction might depend on the vendor platform and the specific system. However, slice-encoding-induced eddy currents are also expected to be present on other vendor platforms, requiring further validation.

Note that the phase equalization method has similarities with a self-navigation approach for 3D multi-shot EPI published by Moeller et al. \cite{moeller_self-navigation_2020}. This method also exploits the repetitiveness of a diffusion sequence with multiple diffusion directions and encoding strengths to derive shot-wise phase navigators. However, in the diffusion case, the phase navigators need to be spatially resolved, resulting in higher computational demands and a more complex phase-fitting procedure. It is probably more sensitive to instabilities compared to the first-order phase model presented. Furthermore, the self-navigation approach by Moeller et al. should also suffer from head rotations, so that an integration with run-time motion correction could improve its stability, resulting in a PEERS method with a more complex phase model.

\hypertarget{discussion-synergy}{%
	\subsection{Synergistic relationship}\label{discussion-synergy}}
The shot-wise phase equalization crucially depends on the comparability between the current shot and its peer, i.e. the reference shot. This is only given if motion correction is applied in run-time before the phase equalization, and thus the raw k-space signal remains unaffected by rotations. Correcting motion retrospectively in the joint multi-shot k-space by regridding \cite{jackson_selection_1991} mixes signals from different shots and, thus, dilutes their phase contributions, which limits the correctability of shot-specific dynamics by phase equalization.
Although phase equalization has been shown to improve tSNR also without servo navigation in-vivo (Fig. \ref{fig:fig7_tsnrExampleInvivo}) and in the phantom (Fig. \ref{fig:fig5_shotPhaseEqualizationExample}), its effectiveness is reduced and it could potentially even reduce signal stability in the presence of motion.

At the same time, phase equalization alleviates frequency precision requirements on the servo navigation method, because the frequency parameters are fine-tuned using the EPI data instead. Thus, the combination of both techniques enables intra-volume frequency corrections for long echo times even with short $3.2\unit{ms}$-navigator acquisitions. The navigator-based frequency and phase correction prior to PEERS is still beneficial, because it keeps the EPI signal phases close to the reference and facilitates the non-linear phase unwrapping, which requires more cautious treatment in Ref. \cite{parker_k-space-based_2022}. At the same time, EPI-based corrections avoid potential phase and frequency mismatches between EPI and navigator sampling, which could arise from their respective transient eddy current and vibration characteristics.

PEERS with servo navigation has been shown to consistently improve tSNR by about $12\%$ in this fMRI study, even without instructed motion, with short $2\unit{sec}$ volume acquisitions and after volume realignment. Without realignment, tSNR increases by $60\%$ (Tab. \ref{tab:tsnrStats_shortNav_latex}). For instructed motion (Fig. \ref{fig:fig8_tsnrMotionInvivo_Subject2}), the tSNR was improved by $16\%$ for servo navigation and by $8\%$ for phase equalization, while together achieving a $29\%$ tSNR increase throughout the brain. 

PEERS can also be combined with other prospective motion correction techniques that maintain the signal correspondence in k-space already during acquisition. In addition, PEERS can also be used to correct scans that involve signal averaging, which is also highly repetitive.

\hypertarget{discussion-frequency-precision}{%
	\subsection{Field precision considerations}\label{discussion-frequency-precision}}
Navigator frequency precision was found to increase approximately linear with sampling duration (Fig. \ref{fig:fig9_longNavigatorExample_rel}A). Longer navigator acquisitions therefore reduce noise in the frequency parameters and stabilize the associated phase corrections at the cost of reduced time efficiency of the navigation. Parameter filtering has also been shown to effectively reduce noise in phase corrections but usually reduces the correction bandwidth at the same time. Relative timing $t_{rel}$ (Eq. \ref{eq:nav_phase_corr_rel}) performs the corrections without prolonged navigator samplings or reduced correction bandwidth; however, the technique omits bulk phases from frequency variations during the volume sampling. This was found to be effective for short $2\unit{sec}$ volume samplings but inappropriate for long volume acquisitions as in gradient echo imaging \cite{riedel_run-time_2025}.

In a previous study \cite{riedel_run-time_2025}, the servo navigation has been extended to run-time gradient shimming by calibrating and updating three more gradient shim parameters in the model. Due to the complexity of the frequency corrections studied here, the model has been restricted to zeroth-order field corrections in this work. First- and higher-order dynamic shimming are subject to future work. Similar to the frequency estimation, uncontrolled noise in the navigator parameter estimates is expected to harm stability for higher-order corrections at long echo times, and field sensitivity is expected to increase with navigator duration.
    
\hypertarget{discussion-future-directions}{%
	\subsection{Future research}\label{discussion-future-directions}}
3D EPI has been shown to offer greater flexibility and sensitivity for high-resolution fMRI, but is more sensitive to physiological noise and scanner system imperfections due to the segmented volume acquisitions \cite{poser_three_2010}. The combination of servo navigation and phase equalization in PEERS reduces the sensitivity of 3D EPI to physiological and sampling-related effects. Therefore, the relative comparison of 2D and 3D fMRI is a focus of future projects.

Non-Cartesian 3D multi-shot trajectories, such as 3D stack-of-spirals \cite{monreal-madrigal_combining_2024}, have been shown to offer efficient alternatives for volume encoding with potential for higher functional sensitivity. For EPI data, the algorithmic extension of phase equalization for arbitrary trajectories provides results comparable to the EPI-specific solution. The performance for non-EPI trajectories remains to be investigated.

In addition, servo navigation and phase equalization have been tested in six healthy and experienced volunteers in this study with a technical focus on enhancements in imaging stability measured by tSNR. However, the impact of these findings on the extraction of functional activity and connectivity from fMRI data requires further investigation. Therefore, future studies involve evaluating these methods with dedicated study designs and data analysis pipelines.




\section{Acknowledgments}
This work has received funding from the European Union’s Horizon 2020 research and innovation program under grant agreement No. 885876 (AROMA project). In addition, the technical and research support from Philips Healthcare is gratefully acknowledged.
%



\bibliography{references} 
\section{Tables}

\begin{table}[H]
    \centering

\caption{Mean tSNR values over all six subjects are given after realignment for different anatomies. The rightmost column shows \emph{Whole brain} before realignment for reference. Abbreviations: GM, gray matter; WM, white matter; CSF, cerebrospinal fluid.}
\begin{tabular}{l|cc|ccccc}
     & Servo & Phase & GM & WM & CSF & Whole brain & Whole brain\\
    Method & navigation & equalization & & & & & not aligned\\\hline
    \csvreader[late after line=\\]%
    {figures/tsnrStats_shortNav_latex.csv}{method=\method,servo=\servo,pe=\pe,gm=\gm,wm=\wm,csf=\csf,whole=\whole,wholeUnaligned=\wholeUnaligned}%
    {\method & \servo & \pe & \gm & \wm & \csf & \whole & \wholeUnaligned}%
\end{tabular}
    \label{tab:tsnrStats_shortNav_latex}
\end{table}

\section{Figures}

\renewcommand{\thefigure}{S\arabic{figure}}

\def\table{\def\figurename{Table}\figure}
\setcounter{figure}{0}
\setcounter{section}{0}
\renewcommand{\thesection}{S\arabic{section}}
\renewcommand\thesubfigure{\roman{subfigure}}

\pagebreak
\thispagestyle{empty}
\section{Supporting Figures}
\pagebreak
\thispagestyle{empty}
\begin{figure}[H]
     \centering
     \begin{subfigure}[b]{\textwidth}
         \centering
         \ifincludeplots\optincludegraphics[width=\textwidth]{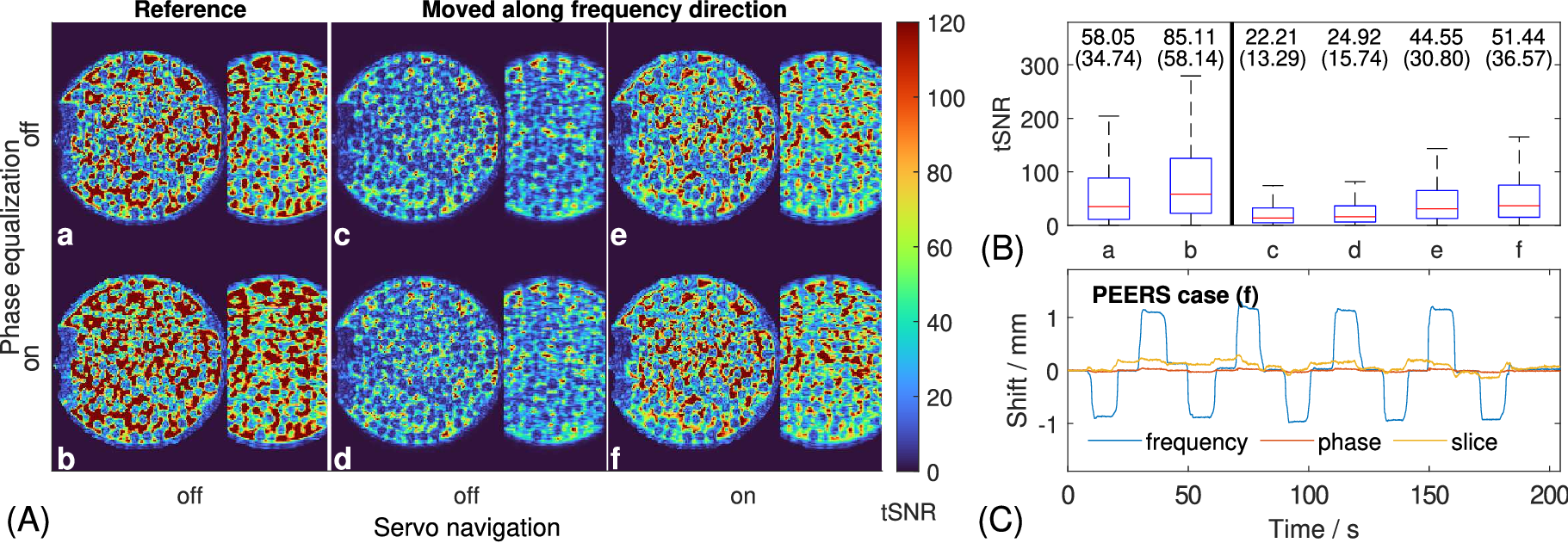}\fi
         \caption{Motion in frequency encoding direction}
         \label{fig:figS1_tsnrMotionPhantom_frequency}
     \end{subfigure}
     \hfill
     \begin{subfigure}[b]{\textwidth}
         \centering
         \ifincludeplots\optincludegraphics[width=\textwidth]{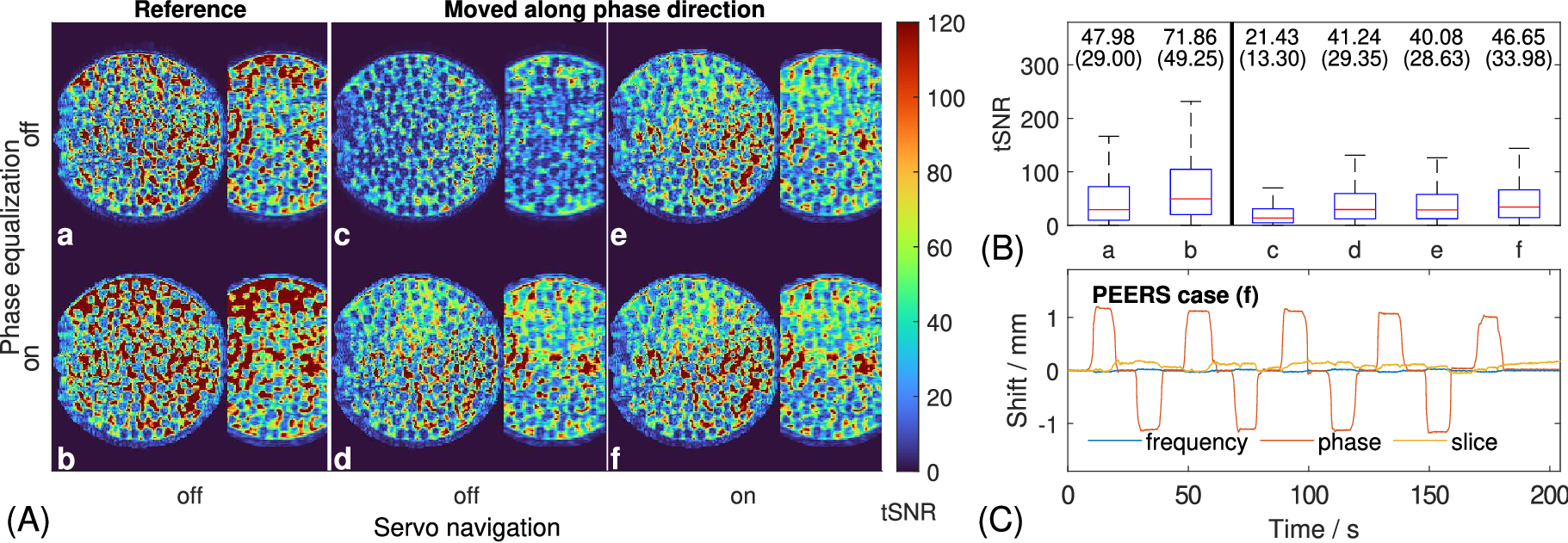}\fi
         \caption{Motion in phase encoding direction}
         \label{fig:figS1_tsnrMotionPhantom_phase}
     \end{subfigure}
     \hfill
     \begin{subfigure}[b]{\textwidth}
         \centering
         \ifincludeplots\optincludegraphics[width=\textwidth]{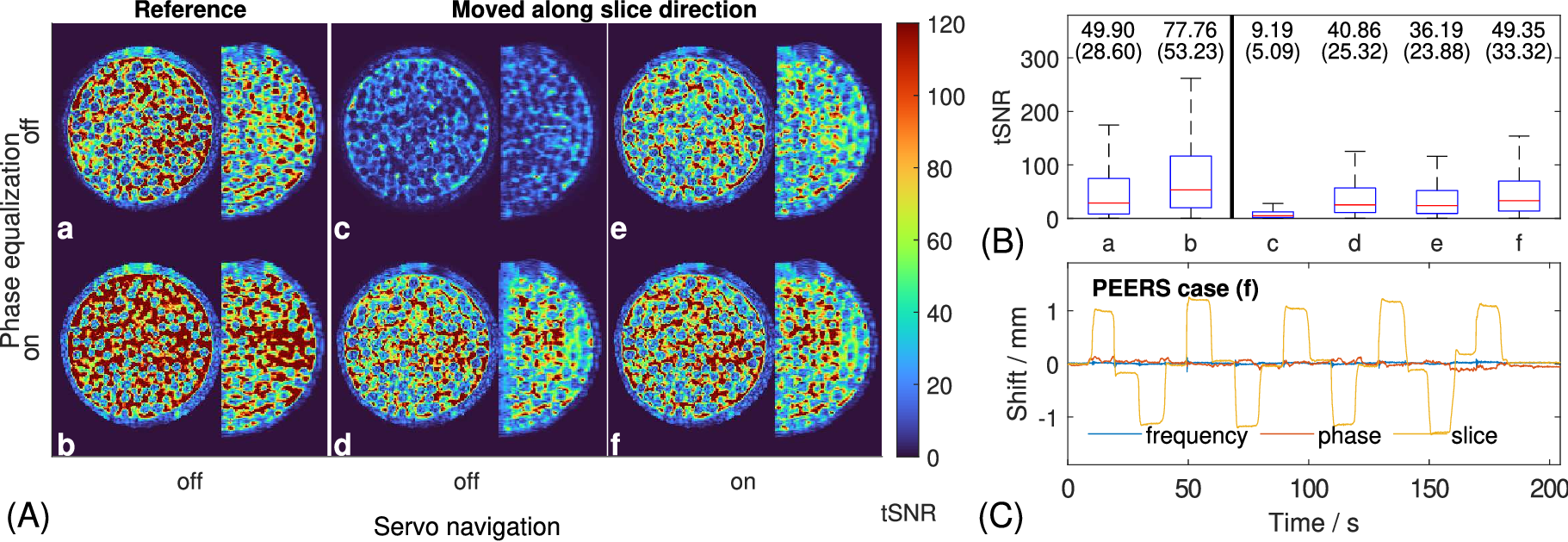}\fi
         \caption{Motion in slice encoding direction}
         \label{fig:figS1_tsnrMotionPhantom_slice}
     \end{subfigure}
        \caption{Phantom scans presented as in Fig. \ref{fig:fig8_tsnrMotionInvivo_Subject2} with intra-scan shifts in (i) frequency, (ii) phase and (iii) slice direction. As shifts in phase and (linear) slice encoding direction appear as a phase ramps and constant phase offsets, respectively, phase equalization is capable of compensating these motions well. Shifts in frequency direction (i) as well as rotations (not performed) cannot be represented in the phase equalization model. Hence, tSNR remains largely uncorrected without servo navigation (d).}
        \label{fig:figS1_tsnrMotionPhantom}
\end{figure}



\end{document}